\documentclass[iop]{emulateapj}
\usepackage{amsmath}
\usepackage{amsfonts}
\usepackage{amssymb}
\usepackage{graphicx}
\usepackage{enumitem}
\begin{document}

\title{\textit{K2} Looks Towards WASP-28 and WASP-151}
\author{T.~Mo\v{c}nik$^{1,2}$, C.~Hellier$^{1}$, and D.~R.~Anderson$^{1}$}
\affil{$^{1}$Astrophysics Group, Keele University, Staffordshire, ST5 5BG, UK\\
$^{2}$Department of Earth and Planetary Sciences, University of California, Riverside, CA 92521, USA}
\email{teom@ucr.edu}
\shorttitle{WASP-28 and WASP-151 Refined by the \textit{K2}}
\shortauthors{Mo\v{c}nik et al.}

\begin{abstract}
By analysing the short-cadence \textit{K2} photometry from the observing Campaign 12 we refine the system parameters of hot Jupiter WASP-28b and hot Saturn WASP-151b. We report the non-detection and corresponding upper limits for transit-timing and transit-duration variations, starspots, rotational and phase-curve modulations and additional transiting planets. We discuss the cause of several background brightening events detected simultaneously in both planetary systems and conclude that they are likely associated with the passage of Mars across the field of view.
\end{abstract}


\section{INTRODUCTION}

Since the failure of a second reaction wheel, the \textit{Kepler} mission was redesigned to observe fields along the ecliptic and renamed into \textit{K2} \citep{Howell14}. Its observing strategy minimized the pointing drift of the spacecraft exerted by the solar radiation pressure, which not only allowed to restore the photometric performance, but also enabled photometric follow-up observations of exoplanetary systems discovered by the ground-based surveys such as WASP \citep{Pollacco06}. The \textit{K2} spacecraft remained operational for 20 scientific observing campaigns with a time-span of around 80 days each before running out of its propellant in October 2018. During the mission lifetime, the \textit{K2} spacecraft provided photometry for a total of 11 previously-discovered WASP planetary systems.

In this paper we present the results from analysing the short-cadence \textit{K2} observations of WASP-28 \citep{Anderson15} and WASP-151 \citep{Demangeon18}, both observed during the Campaign 12.

Many close-in planets, including WASP-28b and WASP-151b, have been found to possess radii well in excess of the standard planet cooling and contraction model \citep{Leconte09}. Knowing precise system parameters is crucial for inferring their bulk composition and their dynamical history, and for understanding what causes the observed radius inflation.

The dearth of short-period Neptune-sized planets, also referred to as the Neptunian desert, has indicated that hot Jupiters and hot super-Earths might have formed and evolved by different mechanisms (e.g. \citealt{Mazeh16}). WASP-151b was found to lie on the upper boundary of the Neptunian desert \citep{Demangeon18}. In this context, it is particularly beneficial to refine parameters for planetary systems that lie close to the Neptunian desert to refine the desert boundaries.

Beside refining the system parameters, we used the \textit{K2} short-cadence data to search for any transit-timing (TTV) and transit-duration variations (TDV), starspot occultations, phase-curve modulations and additional transiting planets. We also provide isochronal age estimates for both systems.

\section{TARGETS}

WASP-28b is an inflated hot Jupiter, transiting an F8 $V=12.0$ star with an orbital period of 3.41 days. The planet was discovered by \citet{Anderson15} using the WASP photometry, two follow-up transit light curves and two sets of radial-velocity measurements obtained by CORALIE \citep{Queloz00} and in-transit HARPS \citep{Mayor03} observations. The system was further photometrically followed-up with ground-based telescopes first by \citet{Petrucci15} who obtained four additional transit light curves  and later by \citet{Maciejewski16} with three additional transits. Both follow-up surveys refined the system parameters and reported the absence of any detectable long-term transit-timing variations.

WASP-151b is an inflated hot Saturn, transiting a G1 $V=12.9$ star with an orbital period of 4.53 days. The discovery of the planet was announced by \citet{Demangeon18}. Beside WASP photometry, they used five follow-up transit light curves from ground-based telescopes and the raw long-cadence \textit{K2} data from the observing Campaign 12. The two accompanying RV datasets were obtained with SOPHIE \citep{Bouchy09} and CORALIE.

\section{\textit{K2} OBSERVATIONS AND DATA REDUCTION}

WASP-28 and WASP-151 were both observed by \textit{K2} in the 1-min short-cadence mode during the observing Campaign 12, which ran between 2016 December 15 and 2017 March 4.

The \textit{K2} light curve for WASP-151 has already been presented in the discovery paper \citep{Demangeon18}, but was generated only from the raw 30-min long-cadence data, which were made publicly available almost immediately after the downlink. The light curves presented in this paper were generated using the short-cadence target pixel files, provided following the official data release. Not only does the increased cadence of observations enable transit modelling with higher precision, but officially released data also undergo Science Operations Center's calibration pipeline \citep{Quintana10}. This level-1 pipeline corrects the data in several ways, such as removing the readout smearing effect, subtracting modelled background, assigning quality flags, etc. Compared to the raw data, the level-1 pipelined target pixel files are used to generate more reliable light curves with lower white noise and to exclude data points which might be affected by cosmic rays or instrumental effects.

We downloaded the short-cadence target pixel files for both systems from the Mikulski Archive for Space Telescopes. We performed the photometric extraction with the PyKE {\scriptsize{KEPEXTRACT}} command \citep{Still12} using a fixed aperture mask of 32 and 30 pixels, centered at WASP-28 and WASP-151, respectively. The pixel mask sizes were chosen by trial and error to extract as much stellar flux as possible while covering fewest background pixels.

The dominant systematic error present in the \textit{K2} data is the sawtooth-shaped flux variations, caused by the pointing drift of the spacecraft and consequent pointing corrections by the thruster firings on roughly 6-h time-scales. To correct for these drift artefacts, we used the modified \textit{K2} Systematic Correction pipeline ({\scriptsize{K2SC}}; \citealt{Aigrain16}), optimised for the \textit{K2} short-cadence data \citep{Mocnik17}. With the removal of drift artefacts, we improved the 1-min photometric precision from 1505 to 370\thinspace ppm for WASP-28 and from 1317 to 641\thinspace ppm for WASP-151.

We did not find any evidence of rotational modulations or any other low-frequency periodic light-curve modulations in the drift-corrected version of the light curves. Any other non-periodic low-frequency light-curve modulations were removed with the PyKE {\scriptsize{KEPFLATTEN}} command, which divided the light curve with the mean of the best-fit second-order polynomials with step and window sizes of 0.3 and 3 days, respectively, 2-$\sigma$ clipping, and 5 iterations.

We present in Figures~1 and 2 the normalized and flattened light curves before and after the drift correction, with all the brightening events from Table~1 removed (see Section~3.1). The 5.3-day loss of data collection around BJD 2457790 occurred during the safe mode, which was likely triggered by the spacecraft's software reset.

\begin{figure}
\includegraphics[width=8.5cm]{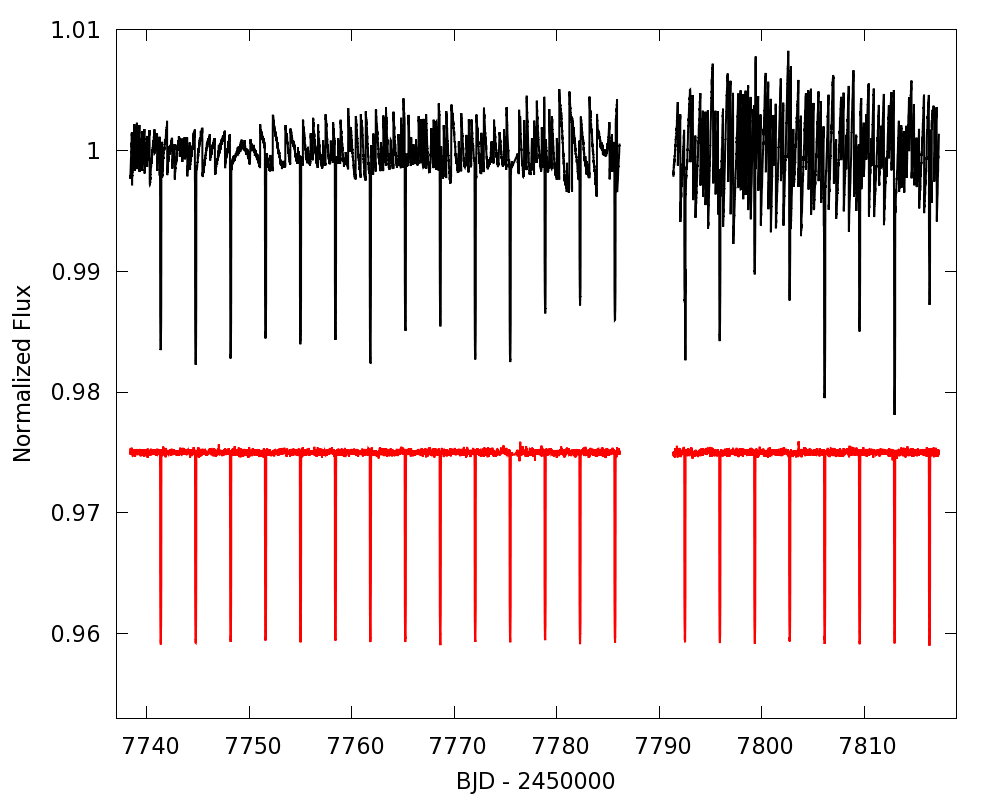}
\caption{Normalized light curve of WASP-28 before (shown in black) and after the drift correction (red), with brightening events and any low-frequency non-periodic modulations removed. The drift-corrected light curve is offset by $-0.025$ for clarity. The observing Campaign 12 covers a total of 22 transits of WASP-28b.}
\end{figure}

\begin{figure}
\includegraphics[width=8.5cm]{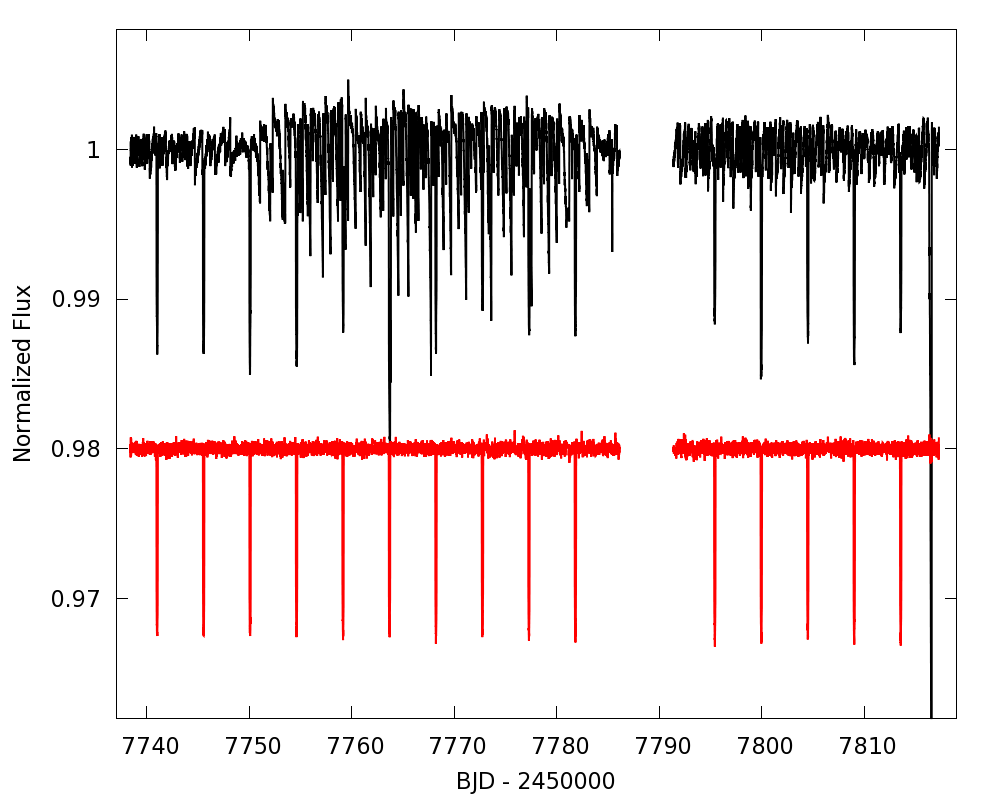}
\caption{Normalized light curve of WASP-151 before (shown in black) and after the drift correction (red), with brightening events and any low-frequency non-periodic modulations removed. The drift-corrected light curve is offset by $-0.02$ for clarity. The observing Campaign 12 covers a total of 15 transits of WASP-151b.}
\end{figure}

WASP-28 was also observed during the Engineering Campaign between 2014 February 4 and 13. We reduced these data in a similar way as for Campaign 12. Because the time stamps in the target pixel files from the Engineering Campaign are not corrected to the solar system barycenter, we applied the barycentric time correction ourselves with an IDL tool {\scriptsize{UTC2BJD}} \citep{Eastman10}. We used the reduced light curve for visual inspection of any potentially interesting events, refinement of the ephemeris, and for the analysis of long-term TTVs. However, we avoided using the light curve for any other aspect of the analysis because the target pixel files from the Engineering Campaign were not fully calibrated by the level-1 pipeline and thus exhibit higher noise and uncertainty which might skew the results.

\subsection{Background Brightening Events}

Upon visual inspection of the corrected light curves we found five episodes where the intensity of the background increased (these are listed in Table~1).

Two uniform background brightening events occurred at BJD 2457770 and 2457774 in both light curves, each lasting for $\sim$1\thinspace h. A similar event affected the WASP-28 background at BJD 2457775 while a 4.8-h-long event affected WASP-151 at BJD 2457781. The longest brightening event lasted nearly 0.5 days, and was non-uniform and seen only in WASP-28 light curve around BJD 2457776.

We believe that these background brightening events may have been caused by the passage of Mars through the field of view. This is surprising because both targets are located far from the areas on the \textit{K2} detector affected by the direct, spilled or ghost images of Mars (WASP-28 was located on CCD module 9-1 and WASP-151 on module 15-4). All the uniform brightenings coincide within one hour with the centre of Mars entering or leaving certain CCD modules (see Table~1). Although this might be a coincidence, it suggests that the uniform brightening events may have been caused by some sort of instrumental effect such as a video crosstalk. We did not find any correlation between Martian CCD module crossings and the longest and non-uniform background brightening event in WASP-28, which may instead have been caused by a (reflected) ghost image of Mars.

In addition to the above background brightening events, the light curve of WASP-151 exhibited another brightening event centered at BJD 2457799. Inspection of individual images revealed that this was caused by an asteroid flyby, whose path crossed the center of WASP-151.

\begin{table}
\centering
\caption{Light-curve brightening events (listed are the first and last BJD of when the event is taking place, relative peak brightening, and a suspected cause for the event)}
\begin{tabular}{lcc}
\hline\hline
BJD $-$ 2457700&$F_{max}$&\textit{cause}\\
\hline
\noalign{\smallskip}
\textbf{WASP-28}\\
70.022 - 70.058&0.0053&Mars enters module 24-3\\
74.312 - 74.401&0.0045&Mars leaves module 24-2\\
75.181 - 75.228&0.0053&Mars enters module 19-3\\
75.650 - 76.140&0.0097&ghost image of Mars (?)\\
\noalign{\smallskip}
\textbf{WASP-151}\\
70.006 - 70.064&0.0081&Mars enters module 24-3\\
74.303 - 74.367&0.0070&Mars leaves module 24-2\\
80.940 - 81.141&0.0063&Mars enters module 13-1\\
99.483 - 99.528&0.0075&asteroid flyby\\
\hline
\end{tabular}
\end{table}

All the light-curve brightening events combined affected a total of 0.66 days (0.9\%) and 0.37 days (0.5\%) of WASP-28 and WASP-151 observations, respectively. We removed the brightening events from the light curves to avoid any potential impact on the following analysis.

\section{SYSTEM PARAMETERS}

We obtained the system parameters with the simultaneous Markov chain Monte Carlo (MCMC) analysis of the transit light curves and radial velocity (RV) measurements. The MCMC procedure is presented in \citet{CollierCameron07}, \citet{Pollacco08}, and \citet{Anderson15}. The photometric input for each system was the normalized and flattened \textit{K2} Campaign 12 light curve from Section~3. The spectroscopic inputs were the publicly available RV measurements published in the corresponding discovery papers, \citet{Anderson15} for WASP-28 and \citet{Demangeon18} for WASP-151. Because the HARPS RV dataset for WASP-28 covered the planetary transit, we were able to fit also the Rossiter-McLaughlin effect. The four-parameter limb-darkening coefficients were interpolated at each MCMC iteration through the tables of \citet{Sing10} and are reported in Tables~2 and 3 as the post-burn-in median values.

The initial values for all MCMC jump parameters, except the stellar mass and eccentricity (see the following two paragraphs), were selected based on the values reported in the corresponding discovery papers. The stellar masses and effective temperatures were constrained with Gaussian priors, while the prior probability distributions of all other parameters were treated as being uniform. The parameter step sizes were set to evolve during the MCMC procedure, ensuring that roughly 25\% of proposal sets were accepted via the Metropolis--Hastings algorithm (see \citet{Pollacco08} and \citet{Anderson15} for details).

The stellar masses were determined with the Bayesian mass and age estimator {\scriptsize{BAGEMASS}} \citep{Maxted15}. {\scriptsize{BAGEMASS}} compares the stellar density, metallicity, and spectroscopic effective temperature to stellar evolution models, and applies the MCMC procedure to find best-fitting isochrone. The stellar densities were derived from the transit observables with the initial MCMC system parameter analyses, whereas the stellar effective temperatures and metallicities were taken from the corresponding discovery papers. The stellar masses obtained with the {\scriptsize{BAGEMASS}} procedure were fed back into the main MCMC system parameter analysis as a two-step iteration process. The derived isochronal age estimates are given along with the system parameters in Tables~2 and 3.

Because the close-in planets are expected to circularise on very short time-scales, we imposed circular orbits in the main MCMC run as suggested by \citet{Anderson12}. In a consequent MCMC run, we then set the eccentricity to be fitted as a free parameter. This allowed us to estimate the eccentricity upper limit from the resulting MCMC chain.

To improve the ephemeris, we performed another MCMC analysis, which employed all the publicly available photometric datasets from the discovery and photometric follow-up papers. We also included the WASP-28 light curve from the \textit{K2} Engineering Campaign from Section~3. As there were several different telescopes with different filters used to obtain these datasets, we sourced the additional limb-darkening coefficients from the appropriate tables of \citet{Claret00} and \citet{Claret04}. This approach increased the observations baseline from 79 days to 8.3 years for WASP-28 and 8.6 years for WASP-151, which improved the orbital period precision by factors of 10.8 and 3.4, respectively.

The final MCMC transit models are shown in Figures~3 and 4, plotted over the measured \textit{K2} Campaign 12 light curves of WASP-28 and WASP-151, respectively. The resulting system parameters are given in Tables~2 and 3, which also serve as a comparison to the previous studies. We find system parameters largely in agreement with the previous results, generally with significantly improved precision.

\begin{figure}
\includegraphics[width=8.5cm]{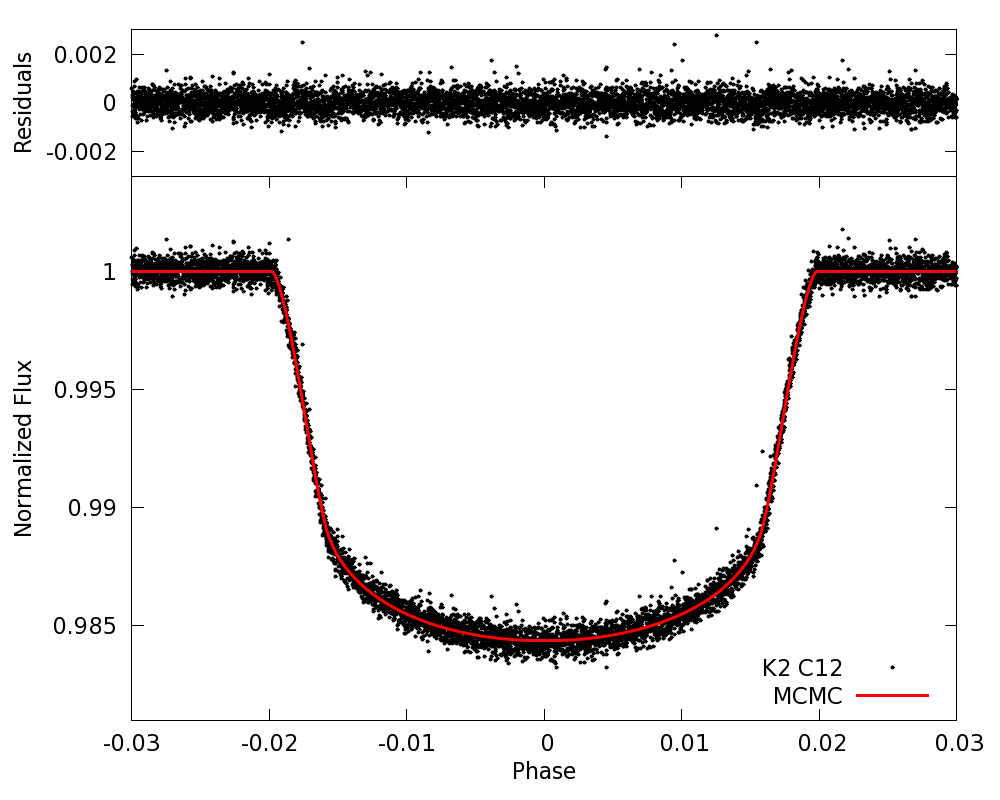}
\caption{Phase-folded light curve of WASP-28 from Campaign 12. The best-fitting MCMC transit model is shown in red. The residuals are shown in the upper panel.}
\end{figure}

\begin{figure}
\includegraphics[width=8.5cm]{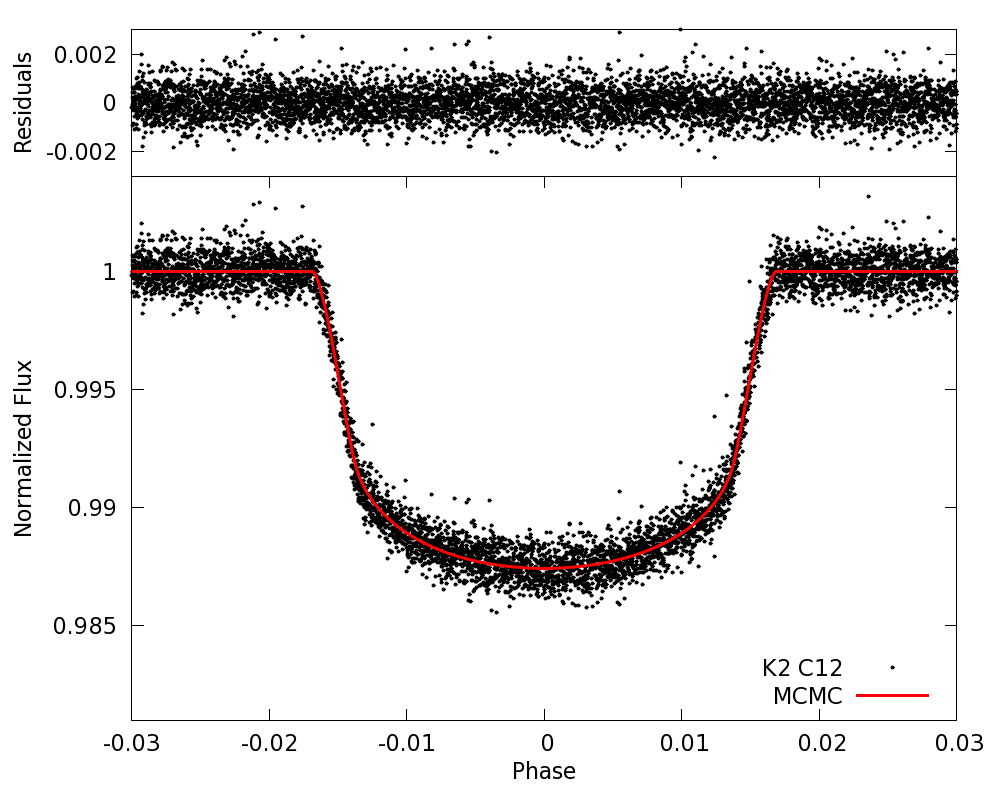}
\caption{Phase-folded light curve of WASP-151 from Campaign 12. The best-fitting MCMC transit model is shown in red. The residuals are shown in the upper panel. The axes scales are the same as in Figure~3.}
\end{figure}

\begin{table*}
\centering
\begin{minipage}{17.5cm}
\caption{MCMC system parameters for WASP-28}
\begin{tabular}{cccccc}
\hline\hline
&this work&\citet{Anderson15}&\citet{Petrucci15}&\citet{Maciejewski16}&\\
Symbol$^{a}$&Value&Value&Value&Value&Unit\\
\hline
$t_{\rm 0}$&$2457642.502126\pm0.000014$&$2455290.40519\pm0.00031$&$2455290.40551\pm0.00102$&$2455290.40595\pm0.00046$&BJD\\
\textit{P}&$3.40883495\pm0.00000015$&$3.4088300\pm0.000006$&$3.408840\pm0.000003$&$3.4088387\pm0.0000016$&d\\
$(R_{\rm p}/R_{\star})^{2}$&$0.013396\pm0.000026$&$0.01300\pm0.00027$&...&...&...\\
$t_{\rm 14}$&$0.13495\pm0.000083$&$0.1349\pm0.0010$&...&...&d\\
$t_{\rm 12}$, \textit{t}$_{\rm 34}$&$0.01469\pm0.00010$&$0.01441\pm0.00070$&...&...&d\\
\textit{b}&$0.228\pm0.013$&$0.21\pm0.10$&...&$0.25^{+0.14}_{-0.25}$&...\\
\textit{i}&$88.514\pm0.090$&$88.61\pm0.67$&...&$88.35^{+1.65}_{-0.92}$&$^{\circ}$\\
\textit{e}&0 (adopted; $<$0.075 at $2\sigma$)&0 (adopted; $<$0.14 at $2\sigma$)&...&...&...\\
\textit{a}&$0.0442\pm0.0010$&$0.04469\pm0.00076$&$0.0445\pm0.0004$&$0.04464\pm0.00073$&au\\
$M_{\star}$&$0.993\pm0.067$&$1.021\pm0.050$&$1.011\pm0.028$&...&M$_\odot$\\
$R_{\star}$&$1.083\pm0.025$&$1.094\pm0.031$&$1.123\pm0.052$&$1.083^{+0.020}_{-0.044}$&R$_\odot$\\
$\rho_{\star}$&$0.7815\pm0.0068$&$0.784\pm0.058$&...&$0.804^{+0.018}_{-0.089}$&$\rho_\odot$\\
$M_{\rm p}$&$0.889\pm0.058$&$0.907\pm0.043$&$0.899\pm0.035$&$0.927\pm0.049$&$M_{\rm Jup}$\\
$R_{\rm p}$&$1.219\pm0.028$&$1.213\pm0.042$&$1.354\pm0.166$&$1.250^{+0.029}_{-0.053}$&$R_{\rm Jup}$\\
$\rho_{\rm p}$&$0.491\pm0.027$&$0.508\pm0.047$&...&$0.474^{+0.041}_{-0.065}$&$\rho_{\rm Jup}$\\
${T_{\rm p}}^{b}$&$1456\pm40$&$1468\pm37$&$1473\pm30$&...&K\\
$\lambda^{c}$&$6\pm17$&$8\pm18$&...&...&$^{\circ}$\\
$\tau_{iso}$&$5.5\pm2.6$&$5^{+3}_{-2}$&$4.2\pm1.0$&...&Gyr\\
$a_{\rm 1}$, $a_{\rm 2}$&$0.386\pm0.004$, $0.602\pm0.017$&...&...&...&...\\
$a_{\rm 3}$, $a_{\rm 4}$&$-0.356\pm0.026$, $0.057\pm0.011$&...&...&...&...\\
\hline
\end{tabular}
\begin{itemize}[leftmargin=0.35cm]
\setlength\itemsep{0cm}
\item[$^{a}$]Meanings of system parameter symbols are given in Table~3.
\item[$^{b}$]Planet equilibrium temperature is based on assumptions of zero Bond albedo and complete heat redistribution.
\item[$^{c}$]Projected spin-orbit misalignment angle was fitted with Hirano model \citep{Hirano11} for the Rossiter-McLaughlin effect.
\end{itemize}
\end{minipage}
\end{table*}

\begin{table*}
\centering
\begin{minipage}{15cm}
\caption{MCMC system parameters for WASP-151}
\begin{tabular}{lcccc}
\hline\hline
&&this work&\citet{Demangeon18}&\\
Parameter&Symbol&Value&Value&Unit\\
\hline
Transit epoch&$t_{\rm 0}$&$2457763.676241\pm0.000040$&$2457741.0081^{+0.0001}_{-0.0002}$&BJD\\
Orbital period&\textit{P}&$4.5334775\pm0.0000023$&$4.533471\pm0.000004$&d\\
Area ratio&$(R_{\rm p}/R_{\star})^{2}$&$0.010664\pm0.000048$&$0.01021^{+0.00010}_{-0.00007}$&...\\
Transit width&$t_{\rm 14}$&$0.15268\pm0.00023$&$0.15250^{+0.00083}_{-0.00042}$&d\\
Ingress and egress duration&$t_{\rm 12}$, \textit{t}$_{\rm 34}$&$0.01564\pm0.00027$&...&d\\
Impact parameter&\textit{b}&$0.304\pm0.023$&...&...\\
Orbital inclination&\textit{i}&$88.25\pm0.14$&$89.2\pm0.6$&$^{\circ}$\\
Orbital eccentricity&\textit{e}&0 (adopted; $<$0.15 at $2\sigma$)&$<$0.003&...\\
Orbital separation&\textit{a}&$0.0550\pm0.00084$&$0.055\pm0.001$&au\\
Stellar mass&$M_{\star}$&$1.081\pm0.049$&$1.077\pm0.081$&M$_\odot$\\
Stellar radius&$R_{\star}$&$1.181\pm0.020$&$1.14\pm0.03$&R$_\odot$\\
Stellar density&$\rho_{\star}$&$0.656\pm0.014$&$0.72^{+0.02}_{-0.04}$&$\rho_\odot$\\
Planet mass&$M_{\rm p}$&$0.316\pm0.031$&$0.31^{+0.04}_{-0.03}$&$M_{\rm Jup}$\\
Planet radius&$R_{\rm p}$&$1.187\pm0.021$&$1.13\pm0.03$&$R_{\rm Jup}$\\
Planet density&$\rho_{\rm p}$&$0.189\pm0.018$&$0.22^{+0.03}_{-0.02}$&$\rho_{\rm Jup}$\\
Planet equilibrium temperature$^{a}$&$T_{\rm p}$&$1312\pm16$&$1290^{+20}_{-10}$&K\\
Isochronal age estimate&$\tau_{iso}$&$5.5\pm1.3$&$5.1\pm1.3$&Gyr\\
\textit{K2} limb-darkening coefficients&$a_{\rm 1}$, $a_{\rm 2}$&$0.599\pm0.005$, $-0.116\pm0.023$&...&...\\
&$a_{\rm 3}$, $a_{\rm 4}$&$0.601\pm0.033$, $-0.336\pm0.013$&...&...\\
\hline
\end{tabular}
\begin{itemize}[leftmargin=0.35cm]
\setlength\itemsep{0cm}
\item[$^{a}$]Planet equilibrium temperature is based on assumptions of zero Bond albedo and complete heat redistribution.
\end{itemize}
\end{minipage}
\end{table*}

\section{NO TTV OR TDV}

The long-term deviations of transit timings from the expected ephemeris can be caused by an orbital decay due to the tidal interactions between a transiting planet and its host star (e.g. \citealt{Murgas14}), whereas periodic TTVs are indicative of the gravitational influence of additional planetary or stellar companions in the system (e.g. \citealt{Holman05}). Similarly, the gravitational interactions can also produce TDVs, albeit at lower amplitudes (e.g. \citealt{Nesvorny13}).

To measure the TTVs and TDVs from the \textit{K2} datasets, we ran an MCMC analysis of system parameters as described in Section~4 on each transit individually by keeping all MCMC jump parameters fixed at values listed in Tables~2 and 3, except the mid-transit times and transit durations. Then, we subtracted the individual transit timings and durations from the ephemeris given in Tables~2 and 3.

\citet{Petrucci15} and \citet{Maciejewski16} have both provided a long-term TTV upper limit for WASP-28 of about 3\thinspace min. Adding our TTV measurements from the \textit{K2} data confirms the absence of long-term TTVs, with a weighted standard deviation of 23\thinspace s and a $\chi_{red}^2$ of 1.32. We show in Figure~5 our TTVs from Campaign 12 and Engineering Campaign (listed in Table~4), along with the measurements from the previous surveys. One TTV measurement around BJD 2456510 from \citet{Petrucci15} and one \textit{K2} measurement near BJD 2457810 have been excluded as $>$3$\sigma$ outliers.

\begin{figure}
\includegraphics[width=8.5cm]{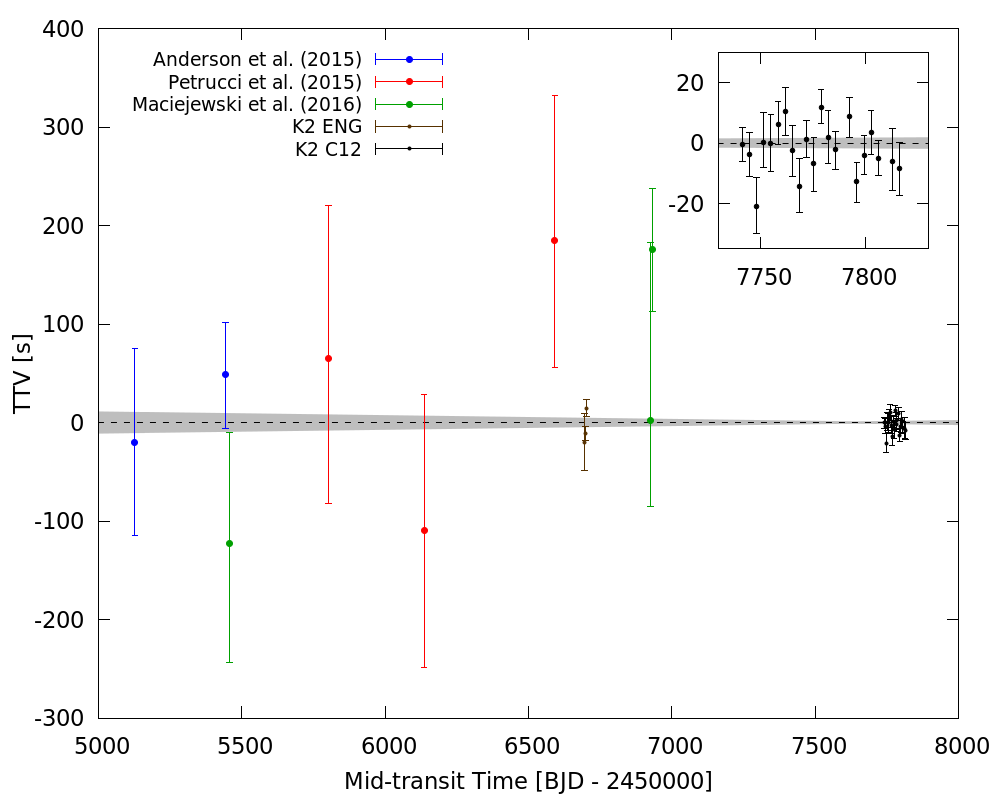}
\caption{The transit-timing stability of WASP-28. The data points clustered around BJD 2456700 and 2457800 correspond to the data taken during the \textit{K2} Engineering Campaign and Campaign 12, respectively. Lighter-coloured data points are TTVs from the various ground-based follow-up surveys, sourced from a collective table published in \citet{Maciejewski16}. The gray area is the ephemeris uncertainty as given in Table~2. The inset plot offers a scaled up view of the TTV measurements from Campaign 12.}
\end{figure}

\begin{table}
\centering
\caption{Transit timings of WASP-28b and WASP-151b from the \textit{K2} Engineering Campaign and Campaign 12}
\begin{tabular}{ccc}
\hline
\multicolumn{2}{c}{BJD -- 2450000}\\
WASP-28b&WASP-151b\\
\hline
$6694.845773\pm0.000335$&$7741.00902\pm0.00019$\\
$6698.254710\pm0.000080$&$7745.54231\pm0.00025$\\
$6701.663845\pm0.000102$&$7750.07617\pm0.00016$\\
$7741.358331\pm0.000065$&$7754.60937\pm0.00021$\\
$7744.767126\pm0.000085$&$7759.14284\pm0.00017$\\
$7748.175763\pm0.000108$&$7763.67619\pm0.00020$\\
$7751.584844\pm0.000105$&$7768.20995\pm0.00016$\\
$7754.993674\pm0.000111$&$7772.74295\pm0.00023$\\
$7758.402583\pm0.000081$&$7777.27675\pm0.00018$\\
$7761.811468\pm0.000092$&$7781.81004\pm0.00021$\\
$7765.220151\pm0.000098$&$7795.41069\pm0.00023$\\
$7768.628849\pm0.000103$&$7799.94422\pm0.00020$\\
$7772.037865\pm0.000072$&$7804.47761\pm0.00021$\\
$7775.446607\pm0.000103$&$7809.01089\pm0.00023$\\
$7778.855658\pm0.000065$&$7813.54409\pm0.00015$\\
$7782.264378\pm0.000100$&\\
$7785.673165\pm0.000073$&\\
$7792.490961\pm0.000077$&\\
$7795.899545\pm0.000078$&\\
$7799.308482\pm0.000075$&\\
$7802.717405\pm0.000085$&\\
$7806.126141\pm0.000067$&\\
$7809.535271\pm0.000076$&\\
$7812.943799\pm0.000118$&\\
$7816.352606\pm0.000103$&\\
\hline
\end{tabular}
\end{table}

We also do not find any statistically significant short-term periodicities or drifts in the TTV or TDV measurements during the Campaign 12 data alone in either of the two planetary systems. With the hypothesis of linear transit timings, the $\chi_{red}^2$ for TTVs were 1.14 for WASP-28 and 1.23 for WASP-151, with semi-amplitude 2-$\sigma$ upper limits of 16\thinspace s and 33\thinspace s for periods shorter than 80 days. For TDV, the assumption of constant transit durations yields the $\chi_{red}^2$ values of 0.54 for WASP-28 and 1.07 for WASP-151, with corresponding semi-amplitude upper limits of 31\thinspace s and 90\thinspace s. The non-detections of periodic TTVs or TDVs render the existence of non-transiting close-in massive planets unlikely. For example, the estimated 2-$\sigma$ upper mass limit of an outer non-transiting planet in a circular orbit within 10\% of the 2:1 resonance is 9.3\thinspace $M_{\rm Earth}$ in WASP-28 system and 14.4\thinspace$M_{\rm Earth}$ in WASP-151, according to relations of \citet{Lithwick12}.

\section{NO STARSPOTS}

If a planet occults a starspot it produces a short in-transit brightening event (e.g. \citealt{Silva03}). Recurring starspot occultation events can be used to derive a stellar rotational period and a constraint on the spin-orbit misalignment angle. Even the non-recurring occultations can in some configurations constrain the misalignment angle if the transit chord passes the active stellar latitudes at only certain transit phases, such as in the case of a misaligned HAT-P-11 system \citep{Sanchis11}.

\citet{Mocnik16a} have demonstrated that the detection of starspot occultation events is possible with two-wheeled \textit{K2} observations, as revealed by the detection of several recurring starspot occultations in the \textit{K2} light curve of an aligned WASP-85 system.

To search for occultation events in the \textit{K2} light curves of WASP-28 and WASP-151 we subtracted the best-fitting MCMC transit models and carefully visually examined the residual transit light curves for in-transit occultation ``bumps''. We found no occultation events with amplitudes above conservative upper limits of 750 and 1000\thinspace ppm, for WASP-28 and WASP-151, respectively.

The non-detection of occultation events is consistent with the spectral types of F8 for WASP-28 and G1 for WASP-151, and is in agreement with the absence of any detectable rotational modulation in both systems.

\section{NO PHASE-CURVE MODULATIONS}

In addition to the transit, the orbit of a planet around its host star can produce a reflectional and thermal modulation, Doppler beaming, ellipsoidal modulation, and secondary eclipse (e.g. \citealt{Esteves13}).  Because a secondary eclipse is a result of a temporarily-blocked reflectional and thermal modulation, its depth is greater or equal to the reflectional modulation amplitude. The planet's thermal emission is expected to be weak in the optical wavelengths of the \textit{K2} telescope \citep{Shporer17}.

To look for such effects we first removed any low-frequency non-periodic variabilities using the {\scriptsize{KEPFLATTEN}} tool \citep{Still12} with window and step sizes of 3 and 0.3 days, for WASP-28, and 5 and 1 day, for WASP-151, respectively. The window and step sizes were chosen as the best compromise between the efficiency of removing the unwanted non-periodic variability and retaining as much phase-curve modulations signal as possible, which we probed with the phase-curve signal injection and recovery tests. Next, we phase-folded the flattened and normalized light curves and binned the phase curves to 200 bins.

We found no phase-curve modulations, and conservatively estimate the semi-amplitude upper limits to be 80 and 90\thinspace ppm, for WASP-28 and WASP-151, respectively. These upper limits take into account the possible partial removal of the phase-curve signal by the light-curve flattening procedure, which we estimated by signal injection and recovery tests. However, we are able to provide tighter semi-amplitude upper limits for the reflectional modulations from the absence of secondary eclipses, because the light-curve flattening procedure did not affect phase-curve signals at such short $\sim$3-hour-long time-scales. These are 40\thinspace ppm for WASP-28 and 80\thinspace ppm for WASP-151.

Using the system parameters from Section~4, the theoretically expected semi-amplitudes for reflectional modulation, Doppler beaming and ellipsoidal modulations are $166A_{\rm g}$\thinspace ppm, 1.6\thinspace ppm and 1.2\thinspace ppm, respectively for WASP-28, where $A_{\rm g}$ is the planet's geometric albedo (see the equations of \citealt{Mazeh10}). For WASP-151 the predicted semi-amplitudes are $88A_{\rm g}$\thinspace ppm, 0.5\thinspace ppm and 0.3\thinspace ppm. We can see that the non-detections of Doppler beamings and ellipsoidal modulations are not surprising, given their small predicted semi-amplitudes. On the other hand, the non-detection of secondary eclipses in WASP-28 system suggests that the planet's albedo is smaller than 0.24. For WASP-151 the planet's albedo remains almost unconstrained with an upper limit of around 0.9.

\section{NO ADDITIONAL TRANSITING PLANETS}

To search for additional transiting planets, we first removed the transits of the known planet in each of the two light curves by replacing the transits' normalized and flattened flux with unity. We then used PyKE's {\scriptsize{KEPBLS}} tool \citep{Still12}, which calculates a box-least-square periodogram as formulated by \citet{Kovacs02}. The periodograms did not reveal any significant periodicity peaks in the period range between 0.5 and 30 days. Using the powers of the highest peaks in the residual periodograms, we estimated the transit depths upper limits to be 160\thinspace ppm for WASP-28 and 220\thinspace ppm for WASP-151.

\section{CONCLUSIONS}

We analysed the short-cadence light curves of WASP-28 and WASP-151 which have been observed by the \textit{K2} during Campaign 12. This work significantly refines the system parameters compared to the previous studies, owing to the short cadence and high photometric precision of the drift-corrected \textit{K2} photometry. We did not detect any TTVs, TDVs, starspots, rotational modulations, phase-curve modulations or any additional transiting planets. Instead, we provide tight upper limits.

The absence of close-in companion planets in both systems is not surprising. Among the 424 currently discovered hot giant planets, defined here as planets with orbital periods shorter than 10 days and radii larger than 0.8\thinspace $R_{\rm Jup}$, there are only 10 systems which contain companion planets (NASA Exoplanet Archive). Of these, only two systems have planetary companions with orbital periods shorter than 100 days, namely WASP-47 \citep{Becker15} and Kepler-730 \citep{Canas19}. The scarcity of hot giant planets with close-in companions has been attributed to the high-eccentricity migration mechanism, a process which would destabilize the orbits of close-in companions \citep{Mustill15}.

We observed several episodes where the \textit{K2} background level is brighter than expected. We associated these with the passage of Mars through the \textit{K2}'s field of view during the Campaign 12, although both targets were positioned on the detector far from any of the areas where the contamination by Martian light was expected. Instead, we think that the brightenings might be caused by some instrumental effect such as a video crosstalk induced by the presence of Mars in the field. This might also be affecting other Campaign 12 datasets.

\acknowledgements{We thank the anonymous referee for their constructive comments that helped to improve this paper. We gratefully acknowledge the financial support from the Science and Technology Facilities Council, under grants ST/J001384/1, ST/M001040/1 and ST/M50354X/1. This paper includes data collected by the \textit{K2} mission. Funding for the \textit{K2} mission was provided by the NASA Science Mission directorate. This work made use of PyKE \citep{Still12}, a software package for the reduction and analysis of \textit{Kepler} data. This open source software project was developed and distributed by the NASA Kepler Guest Observer Office. This research has
made use of the NASA Exoplanet Archive, which is operated
by the California Institute of Technology, under contract with
the National Aeronautics and Space Administration under the
Exoplanet Exploration Program.}

\bibliographystyle{apj}
\bibliography{bibliography}

\end{document}